\documentclass[preprint]{revtex4}

\usepackage{graphicx}
\usepackage{amsmath}
\usepackage{epsfig}

\begin{document}

\title{Millimeter wave spectroscopy of rocks and fluids}

\author{John A. Scales}
\affiliation{Department of Physics, Colorado School of Mines,
Golden, Colorado, USA 80401}
\email{jscales@mines.edu}
\author{Michael Batzle}
\affiliation{Department of Geophysics, Colorado School of Mines,
Golden, Colorado, USA 80401}
\email{mbatzle@mines.edu}

\begin{abstract}
  One region of the electromagnetic spectrum that is relatively
  unexploited for materials characterization is the millimeter wave
  band (frequencies roughly between 40 and 300 GHz).  Millimeter wave
  techniques involve free-space (non-contacting) measurements which
  have a length scale that makes them ideal for characterizing bulk
  properties of multicomponent composites where the scale of
  homogeneity is on the order of millimeters.  Such composites include
  granular materials such as rocks, fluid mixtures, suspensions and
  emulsions.  Here we show measurements
   on partially saturated rocks and an oil/water
  mixture, demonstrating that millimeter wave spectroscopy is a
  sensitive yet rapid measure of changing composition.
\end{abstract}
\maketitle

The millimeter wave band of the electromagnetic (EM) spectrum lies
between radio and optical frequencies.  The two broad classes of
experimental techniques associated with these frequency ranges become
progressively less effective for millimeter waves (MMW).  On the one
hand, conductive cables become attenuative and stray capacitances
build up as frequencies approach a few GHz.  On the other hand,
conventional optical sources have limited power in the THz regime and
far-IR lasers, which use IR pumping of molecular transitions, tend to
be large and complex systems.  Further, direct electronic measurement
of phase is not possible at optical frequencies.  Recent advances in
mode-locked lasers have allowed for small photo-conductive antennas to
be used as THz Hertzian dipoles: a sub-picosecond pulse incident on a
suitable photo-conductor will produce electromagnetic radiation in the
terahertz range.  This technique has received a lot of attention
recently\cite{mittleman} but is not the approach that we will take
here.

Our approach is the use a novel type of 
MMW vector network analyzer (VNA).
A VNA is a four-port instrument in that it measures amplitude and
phase of the transmitted and reflected EM field at each 
frequency.  
A VNA allows one to measure the complete, frequency dependent
transfer function of device or material.

\begin{figure*}
  \centerline{\includegraphics[width=16cm]{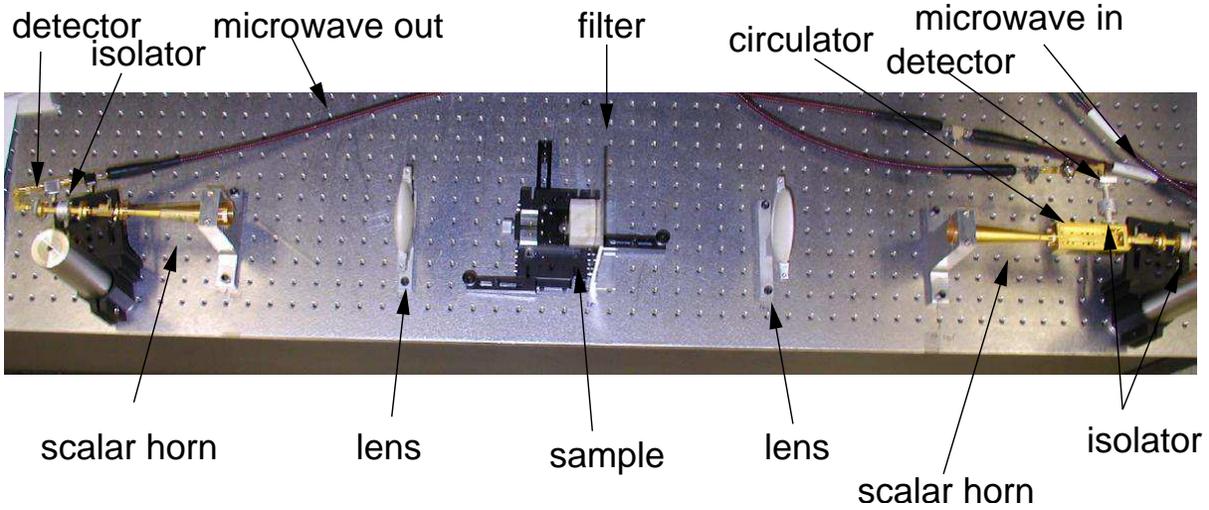}}
  \caption{Quasi-optical setup for the dielectric measurements.
Centimeter waves (8-18 GHz) are created by a sweapable source;
the millimeter waves are then created by a harmonic multiplier (not shown
in the picture).  These MMW are couopled into waveguide and then 
radiated into free space.  The waveguide shown is for the W-band (75-110 GHz)
and the holes on the optical bench are 25 mm apart.  The sample
is placed in the focal plane of the transmitter lens.}
  \label{quasioptics} 
\end{figure*}

At MMW frequencies, the propagation is via free space or waveguide.
Figure \ref{quasioptics} shows a bench-top MMW
setup involving sources/receivers, lenses, circulators, isolators,
transmission through and reflection from small samples.  
Conceptually, this is no different than
an optical measurement, so one uses the 
term quasi-optical for such experiments.
These experiments were performed with the 
MMW VNA developed by AB Millimetre in Paris\cite{abm}.
The millimeter waves are generated by a sweepable centimeter
wave source (i.e., microwaves; in this case from 8-18 GHz), 
not shown in the figure.  
These
centimeter waves are harmonically multiplied by
Schottky diodes, coupled into
waveguide and eventually radiated into free space by
a scalar horn antenna.  A polyethylene lens focuses
the beam  and a sample
is placed
in the focal plane.
The transmitted field is then
collected by an identical lens/horn combination, detected
by another Schottky harmonic detector and fed to a vector receiver
which mixes the centimeter waves down to more easily manageable 
frequencies where the signal is digitized.
Reflected waves are also collected by the transmitting horn and
routed via a circulator and isolator to the vector receiver.
The source and receiver oscillators are phase-locked.
The experiments described here were performed 
in the W band (nominally 75-110 GHz); other bands
are readily accessible by changing waveguides and sources/detectors.
A more complete description of the system is given
in \cite{mmwspectroscopy} and \cite{mvna:rsi}.  

The most
common alternative to solid state devices 
for producing coherent, sweepable 
MMW
is the backward wave oscillator (BWO).
A BWO is a periodically loaded electron
tube, so named because the matching of the phase and beam
velocities gives a negative group velocity.  While potentially
quite powerful, the BWO requires high voltage, temperature and
vacuum for the electron gun, as well as a large permanent
magnet and precise machining of the periodic deceleration
electrodes.
Application of these
sources to millimeter and sub-millimeter wave spectroscopy can be
found in \cite{bwo,kawate}.  For our purposes, solid state sources
producing a few milliwatts of power are perfectly adequate.

If we use samples that are large in the direction transverse to
the beam and with plane parallel faces in the longitudinal 
direction, then the samples become Fabry-Perot cavities (or etalons)
\cite{coherentoptics,ajp}.  This means that inside the sample
standing waves are set up as one sweeps over frequency.  Hence
the frequency response of the transmitted and reflected EM 
waves shows the characteristic Fabry-Perot interference fringes.
If there is no dispersion over the band of interest, 
fitting an analytic Fabry-Perot model to the data
is straightforward.   In fact, this is done by the VNA
itself.  The use of Fabry-Perot interferometry in open resonators
is thoroughly described in \cite{fpi}.

The model depends on the permittivity and
the thickness of the sample; we measure the thickness of the
sample and then fit the permittivity.
This makes the technique spectroscopic
and gives it considerable sensitivity to small changes in the
dielectric properties of the sample \cite{ajp}.

\begin{figure}
  \centerline{\includegraphics[width=12cm]{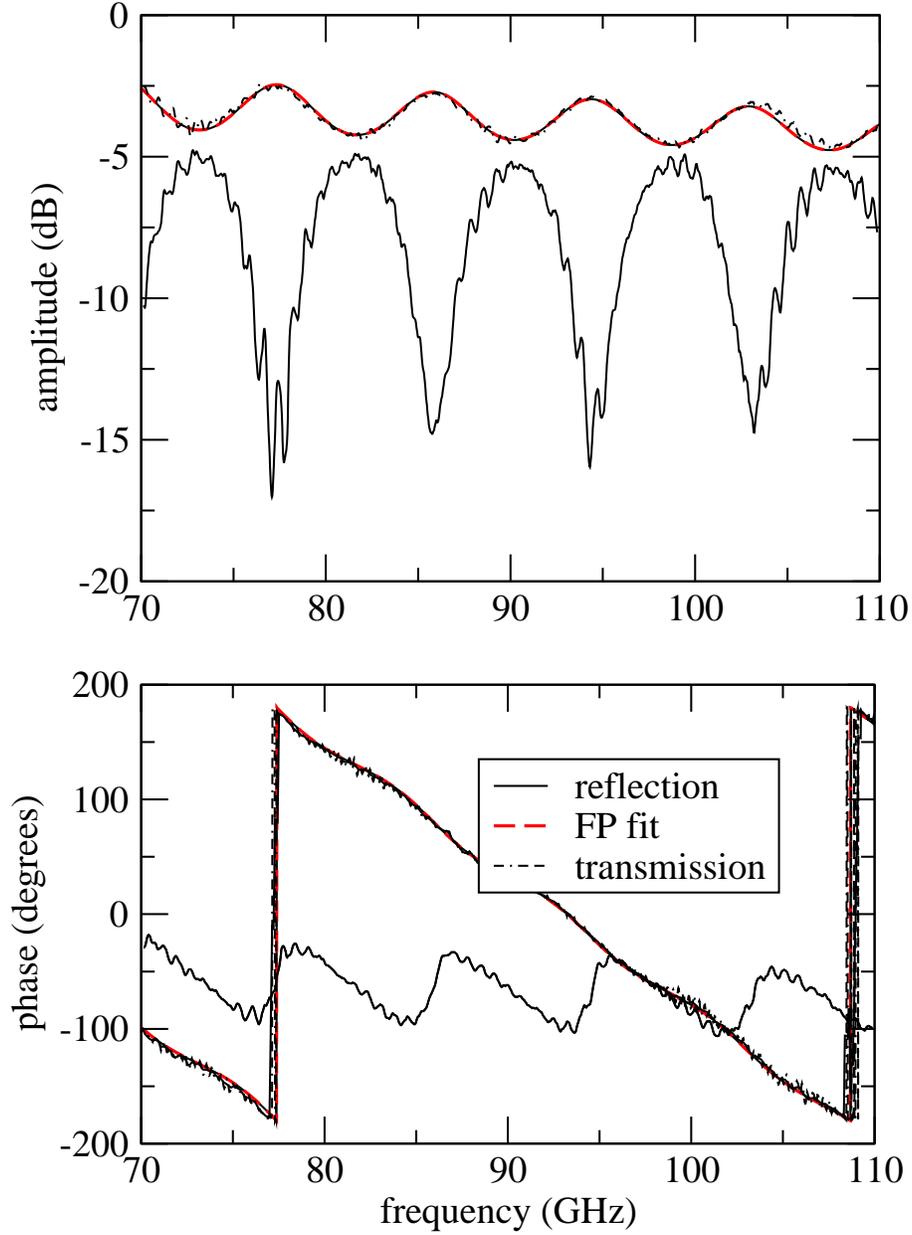}}
  \caption{Amplitude and phase of the transmitted and reflected
electric field as a function of frequency from 75-110 GHz.  
The sample is a 7.1 mm thick piece of standstone.  The
low frequency variations are the Fabry-Perot interference
fringes. The dashed line is the fit to the FP model.}
  \label{emfields}
\end{figure}

For example, Figure \ref{emfields}
shows a sweep from 75 to 110 GHz in a 7.1mm thick sample of 
Berea sandstone.
The top figure shows the amplitude of the transmitted
E field and the bottom shows the phase.  The Fabry-Perot interference
fringes are clear and this allows one to quickly extract the dielectric
constant given the sample thickness. Here we have fit the transmitted
field, but obviously one could also work with the reflected field.

The main advantages of MMW over microwaves are the smaller
wavelengths and shorter time scales.  Smaller wavelengths translate
into smaller antennae (better diffraction properties for a given size)
and higher spatial resolution for heterogeneous samples.
The smaller wavelengths also allows
for easier construction of models of spatially complex systems such
as photonic crystals.
And for dynamical processes (e.g., 
measuring charge mobility), MMW frequencies correspond
to a picosecond time scale rather than the nanosecond scale of microwaves.

Next we show several applications of this MMW
spectroscopy,
both related to the water content of {\it soft} samples.  
This is certainly not a new idea.  Pioneering work in the applications
of millimeter waves to materials characterization has been carried out
for more than 40 years 
in the former Soviet Union.
The paper \cite{bwo} has an interesting history of the
developement and use of the BWO for millimeter
and submillimeter work.  And at the 
Institute of Radio Engineering and Electronics
in Moscow, the group of V. Meriakri has done very interesting work
on the application of these techniques to characterization
of a wide variety practically important materials,
including petroleum
(e.g., \cite{meriakri:oil, meriakri:water}).
Similar techniques were used to assess dielectric measurements
as a means to monitor fluid flow around waste repositories \cite{frasch}.

But advances in both solid state sources/detectors 
and in mode-locked femtosecond lasers make the 
experimental methods more accessible now than ever before.
In particular, our measurements with the VNA allow us to sweep the
entire W-band at several thousand frequencies
in less than a minute and dynamically monitor the slowly changing
properties of the sample.



\begin{figure}
  \centerline{\includegraphics[width=14cm]{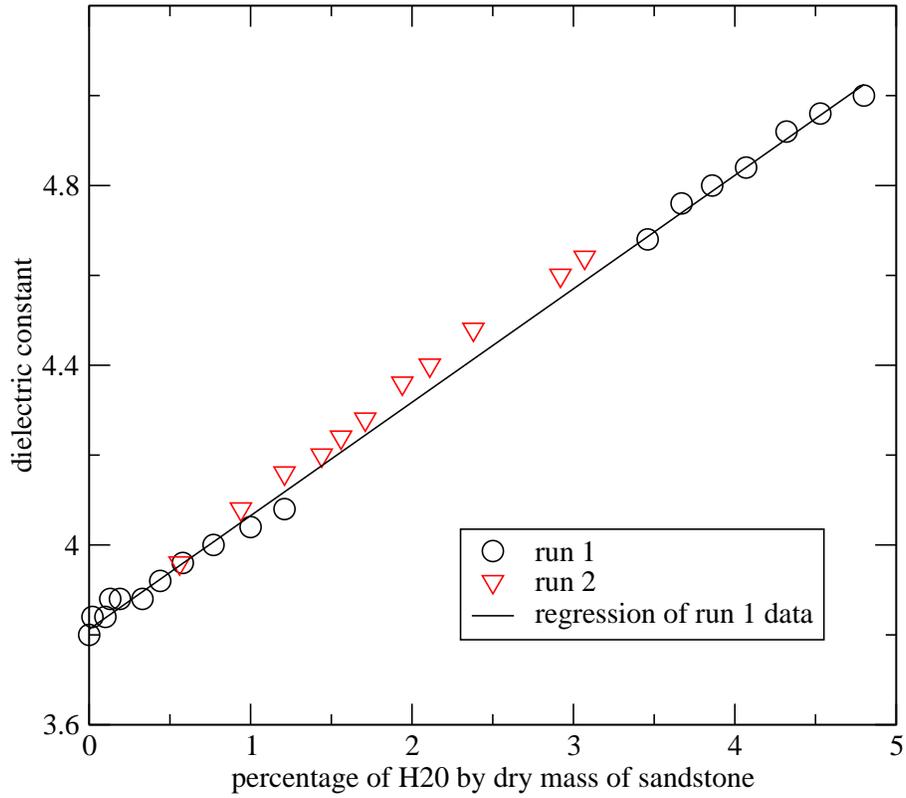}}
  \caption{A 55g sample of Berea stand-stone was partially saturated with
distilled 
water and then allowed to dry out in the lab.  As it dried out,
we measured the dielectric constant by sweeping over 75-110 GHz.  To
check the repeatability of the measurements a second run was performed
24 hours later.  The horizontal axis shows amount of water saturated in
the sample as a fraction of the total dry mass of the sample.}
  \label{dielectric}
\end{figure}

In all these applications we measured the dielectric permittivity as a
function of the water content in the sample. In the first case the
sample was a porous rock (Berea sandstone).  The room-dry sample was
partially saturated by allowing it to sit in distilled water. A 5\%
mass increase due to this simple water imbibition translates into
approximately 50\% saturation of the pore volume (this sandstone has a
porosity of approximately 20\%); or a 10\% fraction of water by total
volume.  Over the course of several hours the dielectric constant and
mass were measured every few minutes, with one break in the middle
(Figure \ref{dielectric}).  The relatively low laboratory humidity
resulted in a decreasing water saturation with time due to evaporation
from the sample surface.  Under ambient 
conditions it took about two hours for the rock to dry out
to its original state.
The measurement was repeated 24 hours later to check the
reproducibility.  Note that the dielectric constant of liquid
water is much lower at millimeter wave frequencies than at
RF.  For example, at 100 GHz $\epsilon$ is 7.72 at 
20 C and 8.16 at 30 C.\cite{liebe}

The repeatability of the permittivity shown in Figure 3 suggests an
experimental uncertainty of less than a percent. This means that by
calibrating a known rock in this way we should be able to use the
permittivity to infer the fluid saturation. The speed of the technique
means that we can study dynamical processes such as evaporation
and flow.

In order to apply this technique to fluids we used Spectrocil
quartz cells from Starna.  These had an interior thickness of 9.5
mm along the beam direction
and where much larger than the beam diameter.   The calibration
techniques associated with the VNA we use allow one to make
a baseline measurement with no sample present.
This calibrates out the instrument response.
But in addition, we used this capability
to calibrate out the presence of an
empty cell.  Then we make sweeps over the W band with the cell filled
with oil.  This gives us the response of just the fluid in the cell.
For an oil sample with no water present 
we get an index
of 1.52 (permittivity 2.31).  This agrees with published 
values at 2 mm wavelength  \cite{meriakri:water}.

To see how sensitive our technique is to the presence of water,
we prepared two additional samples, by mixing distilled water
with the oil and agitating to form a fine suspension.
One sample had 5\% water and one 10\%.  For these we got permittivities
of, respectivley, 2.43 and 2.66.  Over the course of several
weeks, permittivity meaurements were
repeatable to better than half a percent.
These results are consistent with those in \cite{meriakri:water}
who looked at oil/water emulsions with water content
of .5\% and 1\%.  Both of these sets of measurements are
well-fit by the Maxwell-Garnett
mixing model\cite{maxwell}:
$$
\epsilon = \epsilon_2 \left[
1 +
\frac{3 W (\epsilon_1 - \epsilon_2)}{\epsilon_1 +
2 \epsilon_2 - W(\epsilon_1 -\epsilon_2)}
\right]
$$
where $\epsilon_1$ and $\epsilon_2$ are the permittivities of the
water and oil, respectively and $W$ is the fraction of water.

For example, 
this model predicts 5\% and 10\% water/oil emulsion permittivities
of 2.46 and 2.62 respectively, assuming a water permittivity
at 100 GHz of 7.72\cite{liebe}.

Thus adding 5\% water to the oil
results in an easily measurable
increase in the dielectric constant.  Hence by
calibrating the samples beforehand for a range of water saturations,
it may be possible to use the bulk permittivity to deduce the water
of unknown samples. 
These results are consistent with those measured on Basalt
in \cite{frasch}.

Finally we looked at three samples of oil shale taken from the same
well location but with different composition and, evidently,
permeability.  After measuring the permittivity of the dry
samples, these
were weighed, placed in a vacuum chamber and
vacuum applied for 1 hour.  Then distilled water was injected
to 6.9 MPa pressure (without reexposure to air) and
allowed to soak overnight, then reweighed.

Samples 1 (dry $\epsilon= 5.14$)
and 2 (dry $\epsilon = 5.89$)
saturated to 9.7\% and 8.1\%
by mass, respectively, and showed slightly more than a 2\% increase
in their dielectric permittivity.  On the other hand, sample 3 
(dry $\epsilon=4.96$) took on only
.4\% water and showed a correspondingly small increase in
permittivity of .8\%.  While these results indicate an approximately
linear relation between water content and permittivity, the Maxwell-Garnett
does a poor job in this case, predicting much smaller changes in 
permittivity.   But since this mixing model neglects the
chemical interaction of the water with the rock matrix, this may
not be too surprising for shales, which have significant clay content.

In conclusion, we have shown how use of a state-of-the-art
millimeter wave vector network analyzer allows one
to quickly perform
dielectric spectroscopy on soft materials
such as rocks and fluids.
The technique has many applications in the study of composite mixtures.
The short wavelength gives good spatial resolution, while the
speed of the measurement allows one to observe dynamical processes
such as evaporation and capilary motion of fluids.

\begin{acknowledgments}
We are grateful for technical assistance and discussions with 
Philippe Goy of the Laboratoire Kastler Brossel, 
Ecole Normale Superieure and AB Millimetre in Paris.
This work was partially supported by the National Science
Foundation (EAR-041292).
\end{acknowledgments}

\clearpage

\section*{Figure Captions}

Figure 1: Quasi-optical setup for the dielectric measurements.
Centimeter waves (8-18 GHz) are created by a sweapable source;
the millimeter waves are then created by a harmonic multiplier (not shown
in the picture).  These MMW are couopled into waveguide and then 
radiated into free space.  The waveguide shown is for the W-band (75-110 GHz)
and the holes on the optical bench are 25 mm apart.  The sample
is placed in the focal plane of the transmitter lens.

Figure 2:  Amplitude and phase of the transmitted and reflected
electric field as a function of frequency from 75-110 GHz.  
The sample is a 7.1 mm thick piece of standstone.  The
low frequency variations are the Fabry-Perot interference
fringes. The dashed line is the fit to the FP model.

Figure 3:  A 55g sample of Berea stand-stone was partially saturated with
distilled 
water and then allowed to dry out in the lab.  As it dried out,
we measured the dielectric constant by sweeping over 75-110 GHz.  To
check the repeatability of the measurements a second run was performed
24 hours later.  The horizontal axis shows amount of water saturated in
the sample as a fraction of the total dry mass of the sample.

\clearpage

\end{document}